\documentclass[twocolumn,prb,aps,fleqn,floatfix,amssymb]{revtex4}

\usepackage{amsfonts,bm,amsmath}

\usepackage{graphicx}

\newcommand*{\qq}{\bm{q}}
\newcommand*{\rr}{\bm{r}}
\newcommand*{\xx}{\bm{\xi}}

\newcommand*{\matel}{
h_{\nu_{1}\nu_{2}\nu_{3}\nu_{4}}^{(\sigma_{1}\sigma_{2}\lambda\sigma_{4})}}

\begin{document}
\title{Coulomb matrix elements for the impact ionization process in 
nanocrystals: the envelope function approach} 
\author{Piotr Kowalski, {\L}ukasz Marcinowski, Pawe{\l} Machnikowski}
\date{\today}
\affiliation{Institute of Physics, Wroc{\l}aw University of Technology, 
50-370 Wroc{\l}aw, Poland}
\begin{abstract}
We propose a method for calculating Coulomb
matrix elements between exciton and biexciton states in semiconductor
nanocrystals based on the envelope function formalism. We show
that such a calculation requires proper treatment of the Bloch parts
of the carrier wave functions which, in the leading order, leads to
spin selection rules identical to those holding for optical interband
transitions. Compared to the usual (intraband) Coulomb couplings, the
resulting matrix elements are additionally  
scaled by the ratio of the lattice constant to the nanocrystal radius. As a result,
the Coulomb coupling between exciton and biexciton states scale as
$1/R^{2}$. 
We present also some statistical estimates of
the distribution of the coupling magnitudes and energies
of the coupled states The number of biexciton
states coupled to exciton states form a certain energy range shows a
power-law scaling with the ratio of the coupling magnitude to the energy
separation. We estimate also the degree of mixing between exciton and biexciton
states. The amount of biexciton admixture to exciton states at least 1~eV above
the multiple exciton generation threshold can reach 80\% but varies
strongly with the nanocrystal size. 
\end{abstract}

\maketitle

\section{Introduction}
Limitations of the efficiency of the existing solar cells motivate 
continuous search for new technological solutions that might lead to 
more efficient photovoltaic conversion. One of the fundamental 
limitations on the efficiency of the existing solar cells results from 
the fact that photons with energies higher than the energy gap of the 
semiconductor excite high energetic electrons in its conduction band. 
The excess energy of these charge carriers is dissipated in a phonon 
relaxation processes and is therefore lost for photovoltaic conversion.

\begin{figure}[tb]
\includegraphics[width=60mm]{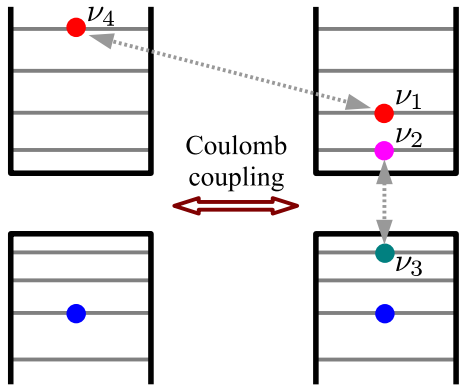}
\caption{Schematic representation of the impact ionization process in a 
nanocrystal. The state labels refer to labeling of the matrix elements
derived in Sec.~\ref{sec:model}.}
\label{fig:ImpactIonization}
\end{figure}

Among systems that are investigated for possible use in solar energy 
conversion devices in order to overcome these losses are semiconductor 
nanocrystals (NCs). The increased efficiency of photovoltaic
conversion in NC-based devices might result from multiple exciton
generation (MEG) due  
to impact ionization (inverse Auger) processes\cite{nozik02}. Such an 
effect consists in generation of two or more electron-hole pairs by a 
single high energy photon and thus converts the excess above-bandgap 
energy into useful current. This process is enabled by Coulomb
coupling between single-pair (exciton, X) states and two-pair
(biexciton, BX)
states in a NC (or, in general, between states with $n$ and
$n+1$ pairs), as shown in Fig.~\ref{fig:ImpactIonization}. It can
result from the system dynamics after optical creation of a single
electron-hole pair, during which an electron relaxes within the
conduction band transferring its energy to an intraband excitation (an
inverse Auger process, in which the two situations in
Fig.~\ref{fig:ImpactIonization} are understood as ``initial'' and
``final''). However, very short time scales of the biexciton generation 
\cite{schaller05} suggest that the process may be instantaneous and
result from the mixing between the X and BX states. In
this case, the original X states contain on the average more than one
electron-hole pair, while the initially dark BX states become
optically active. The MEG process can be thus understood as an
excitation of a BX state mediated by a ``virtual'' X state to which it is
coupled \cite{schaller05} (the two states in
Fig.~\ref{fig:ImpactIonization} are then interpreted as the
``virtual'' state that is directly coupled to light and the final BX
state that is optically excited via the Coulomb-induced
mixing). Clearly, in this picture, the degree of 
mixing between the states and their spectral distributions (spectral
densities) are crucial for the efficiency of the MEG process. Another
possible process \cite{rupasov07} involves Coulomb coupling between
empty NC and BX states and relies on an intermediate BX state.

The MEG effect has been observed using a
variety of experimental techniques in NCs formed of various
narrow-gap  semiconductors and under various excitation conditions
\cite{schaller04,schaller05,ellingson05,schaller07,pijpers07,beard07,ji09},
as well as in systems of coupled NCs \cite{trinh12}.
In some experiments, quantum efficiencies as high as 700\% were observed
\cite{schaller06}. On the other hand, some later experimental
investigation showed much lower efficiencies of the MEG process or even
no traces of MEG at all
\cite{nair07,trinh08,pijpers08,ben-lulu08,nair08}. The subsequent
discussion of the experimental factors involved in extracting the MEG
efficiency from experimental data \cite{mcguire08,mcguire10,binks11}
showed that performing reliable experiments and correctly interpreting
their results is not quite trivial. Uncontrolled effects present 
in the experiments, like photocharging of the NC core and 
charge trapping at the surface ligands, can indeed lead
to a considerably overestimated result \cite{binks11}. In any case,
however, enhanced photocurrent due to MEG has been observed in real 
NC-based solar cells\cite{sambur10,semonin11} providing a
direct proof of  
the usefulness of this process in solar energy conversion. Still, the
experimental results and the controversies they arouse motivate
theoretical work aimed at full understanding of the MEG process in
NCs.  

Theoretical description of the X and BX spectrum and the X-BX
couplings is sought within various approaches available for modeling
semiconductor nanostructures. Atomistic approaches to this
problem include density functional theory \cite{deuk11,deuk12},
pseudopotential 
\cite{franceschetti06,rabani08,califano09,baer12} and 
tight binding \cite{allan06,delerue10,korkusinski10} methods. However,
the high computational complexity of atomistic methods limits the size
of tractable systems and forces one to radically truncate the basis of
single particle functions when computing the properties of
few-particle complexes. Therefore, one often resorts to an envelope
function approach based on the k$\cdot$p theory
\cite{shabaev06,witzel10,silvestri10}, which was very useful in the
past for understanding the fundamental electronic properties of
NCs \cite{efros98}. Apart from establishing the spectral
distribution of the X and BX levels, the central point of any theoretical
modeling of the MEG process is the evaluation of the X-BX Coulomb
matrix element which can then be used, e.g, to find the spectral
properties of correlated X-BX states \cite{korkusinski11} or to
study the system kinetics
\cite{shabaev06,rabani10,witzel10,piryatinski10,velizhanin11a,korkusinski11}. 

In the present paper, we present the calculations of the Coulomb
matrix elements between X and BX states within an envelope function
approach. This approach is much more explicit than the atomistic
computations and yields relatively transparent formulas for the matrix
elements that offer much more insight into the properties of the X-BX
couplings. This allows us to point out that the correct treatment of the Bloch parts
of the carrier wave functions (which was not made explicit in
previous works) leads to selection rules for the newly
created pair identical to those holding for optical transitions. In
addition, the interband character of the coupling reduces the matrix
element (as compared to the usual electron-electron interaction) by a
factor $a/R$, where $a$ is on the order of the lattice 
constant and $R$ is the radius of the NC, and leads to the
$1/R^{2}$ dependence on the NC size. 
On the practical side, the envelope function approach leads to formulas that, for a
spherical NC, can be evaluated at very modest computational
cost. This allows us to find coupled pairs of X-BX states in a very
broad energy window and to study the distribution of the magnitudes of
the matrix elements vs. the energies of the coupled
configurations.
Here, we restrict ourselves to the simplest example of
an application of the proposed formalism and calculate the matrix
elements using a very simple model of wave functions in order to
extract very general statistics on the X-BX couplings. The important
conclusion from this study is that the X-BX state mixing is on the
average dominated by relatively strongly coupled pairs of
energetically close states. This means, on one hand, that admixture of
the BX states to the X states (or conversely) cannot be treated
perturbatively. On the other hand, it shows that truncating the set of
states to an energy window around a given X (or BX) state is a
well-defined and convergent approximation. Quantitatively, the matrix
elements found for an InAs NC are in the meV range, which is
clearly very small compared to typical separations of single particle
states in these structures. Still, admixture of BX states to a given X
state can reach several to a few tens per cent due to relatively large
density of coupled BX states.

The paper is organized as follows. In Sec.~\ref{sec:model}, we derive
general formulas for the Coulomb matrix element between X and BX
configurations in a spherical NC. Next, in
Sec.~\ref{sec:simple-model} we present a simple example of an application of
this formalism. Sec.~\ref{sec:results} contains results on the
statistical distribution of the X-BX couplings and an estimate of the
degree of X-BX mixing. Sec.~\ref{sec: concl} contains concluding
discussion of the results and the outlook for their extension.

\section{Model and general description}
\label{sec:model}

In this section we present a general derivation of the matrix element 
underlying the MEG process in a NC. The calculation will be
performed in the  
envelope function approximation. To simplify the notation we assume 
identical envelopes for all bands but generalization to arbitrary 
envelope wave functions is straightforward. 

\subsection{General considerations}

We calculate the matrix 
elements between single-subband wave functions. The result can be 
generalized to derive the Coulomb couplings between states in more 
accurate models including band mixing. We consider a spherical 
NC of radius $R$ with the dielectric constant 
$\epsilon_{\mathrm{s}}$, emerged in a host material with the dielectric 
constant $\epsilon_{\mathrm{h}}$. The electron-trion coupling 
responsible for the impact ionization process results from the Coulomb 
coupling between the states in conduction and valence bands, 
corresponding to an intraband transition of an electron accompanied by 
the creation of an additional electron-hole pair 
(inverse Auger process) as shown schematically in 
Fig.\ref{fig:ImpactIonization}. The relevant part of the Coulomb 
Hamiltonian is 
\begin{displaymath}
H =
\sum_{\nu_1\nu_2\nu_3\nu_4}\sum_{\lambda\sigma_1\sigma_2\sigma_4}
\matel
a_{\sigma_1\nu_1}^{\dag}
a_{\sigma_2\nu_2}^{\dag}
a_{\lambda\nu_3}
a_{\sigma_4\nu_4}+\mathrm{H.c.},
\end{displaymath}
where $a_{\sigma_{i}\nu_{i}},a^{\dag}_{\sigma_{i}\nu_{i}}$ are the 
annihilation and creation operators for an electron in the conduction 
subband $\sigma_{i}$ and in the envelope state $\nu_{i}$, while
$a_{\lambda\nu_{i}},a^{\dag}_{\lambda\nu_{i}}$ are the annihilation and 
creation operators for an electron in the valence subband $\lambda$ and 
in the envelope state $\nu$. The index $\nu_{i}$ represents the whole 
set of relevant quantum numbers. The graphical interpretation of this 
term is given in Fig. \ref{fig:ImpactIonization}. We write the wave 
functions in the single-band envelope approximation in the form 
$\Psi_{\sigma\nu}(\rr,s)=\psi_{\nu}(\rr)u_{\sigma}(\rr,s)$, where 
$\psi_{\nu}(\rr)$ is the envelope and $u_{\sigma}(\rr,s)$ is the
lattice-periodic Bloch part ($s$ denotes the value of the electron 
spin). The position vector is written as $\rr=\rr_{i}+\xx$, where 
$\rr_{i}$ denotes the center of the $i$th unit cell (u.c.) and $\xx$ 
lies in the first u.c. The integration over the whole NC 
volume is then expressed as a summation over all the unit cells and
integration over one cell. We assume that the envelope functions change 
slowly, so that $\psi_{\nu}(\rr_{i}+\xx)\approx\psi_{\nu}(\rr_{i})$ and 
use the periodicity of the Bloch functions 
$u_{\sigma}(\rr_{i}+\xx)=u_{\sigma}(\xx)$. The matrix element is then 
given by
\begin{eqnarray}
\lefteqn{ 
\matel =  
\sum_{ss'}\sum_{ij}
\int_{\mathrm{u.c.}} d^3\xx \int_{\mathrm{u.c.}} d^3\xx'
} \label{matel1} \\ 
&& \times  
\psi_{\nu_1}^{*}(\rr_{j}) u_{\sigma_1}^{*}(\xx',s')
\psi_{\nu_2}^{*}(\rr_{i}) u_{\sigma_2}^{*}(\xx,s)
U(\rr_{i}+\xx,\rr_{j}+\xx')
\nonumber \\ 
&& \times 
\psi_{\nu_3}(\rr_{i}) u_{\lambda}(\xx,s)
\psi_{\nu_4}(\rr_{j}) u_{\sigma_4}(\xx',s').
\nonumber
\end{eqnarray}

The two-particle interaction energy in a spherical NC is 
composed of the direct Coulomb interaction and the coupling via surface 
polarization due to dielectric discontinuity between the NC 
and the environment \cite{Brus84}, 
$U(\bm{r},\bm{r}')=U_{\mathrm{direct}}(\bm{r},\bm{r}')+U_{\mathrm{pol}}
(\bm{r},\bm{r}')$, 
where 
\begin{eqnarray*}
U_{\mathrm{direct}}(\bm{r},\bm{r}') & = & 
\frac{e^2}{4\pi \epsilon_0 \epsilon_{\mathrm{s}}}
\frac{1}{|\bm{r}-\bm{r}'|}, 
\\ 
U_{\mathrm{pol}}(\rr,\rr') & = & 
-\frac{e^2}{4\pi \epsilon_0 \epsilon_{\mathrm{s}}}\sum_{k}\alpha_{k}
\frac{(rr')^{k}}{R^{2k+1}}P_{k}(\cos\gamma).
\end{eqnarray*} 
Here $\cos \gamma = \rr \cdot \rr'/ \left( rr' \right) $, $P_{k}$ are 
Legendre polynomials,
$\chi_{k}=(k+1)(\epsilon-1)/(k\epsilon+k+1)$
with $\epsilon=\epsilon_{\mathrm{s}}/\epsilon_{\mathrm{h}}$, and we 
take into account only the two-particle part of the surface 
polarization term.

In consequence, the Coulomb matrix element splits into the
corresponding two contributions
\begin{eqnarray}\label{mat-el-total}
h_{\nu_{1}\nu_{2}\nu_{3}\nu_{4}}^{
(\sigma_{1}\sigma_{2}\lambda\sigma_{4})}
= h_{\nu_{1}\nu_{2}\nu_{3}\nu_{4}}^{
(\sigma_{1}\sigma_{2}\lambda\sigma_{4},\mathrm{dir})}
+ h_{\nu_{1}\nu_{2}\nu_{3}\nu_{4}}^{
(\sigma_{1}\sigma_{2}\lambda\sigma_{4},\mathrm{pol})}.
\end{eqnarray}

\subsection{Direct Coulomb coupling}

For the direct Coulomb term, which is singular at $\rr=\rr'$, we split 
the summation into $i=j$ (short-range contribution) and $i\neq j$ (long 
range contribution). For the former, we use the expansion\cite{jackson98} 
\begin{equation}
\frac{1}{|\xx-\xx'|}=\sum_{l=0}^{\infty}
\frac{\xi_{<}^{l}}{\xi_{>}^{l+1}}
\sum_{m=-l}^{l}
\frac{4\pi}{2l+1}
Y_{lm}^{*}(\theta'\phi')Y_{lm}(\theta,\phi),
\label{sph-ham-add}
\end{equation}
where $\xi_{<}=\min(\xi,\xi')$, $\xi_{>}=\max(\xi,\xi')$, $\theta,\phi$ 
and $\theta,\phi'$ are the spherical coordinates of the vectors $\xx$ 
and $\xx'$, respectively, and $Y_{lm}(\theta,\phi)$ are spherical 
harmonics. Since both $\sigma_{1}$ and $\sigma_{4}$ correspond to 
$s$-type conduction band states only the term with $l=0$ is non-zero in 
the integral over $\xx'$ when Eq.~\eqref{sph-ham-add} is substituted to 
Eq.~\eqref{matel1}. This term, however, yields a vanishing integral 
over $\xx$, as $\lambda$ and $\sigma_{3}$ correspond to states with 
$p$ and $s$ symmetries, respectively. Hence, the short-range 
contribution vanishes.

In the long range term, we expand $|\rr_{i}+\xx-\rr_{j}-\xx'|^{-1}$ to 
the linear order in $(\xx-\xx')$. The summation over $i,j$ in 
Eq.~\eqref{matel1} is then replaced by integration, where one formally 
excludes a small volume around $\rr=\rr'$ (which is represented by a 
prime over an integral)\cite{Takagahara93}. Thus one finds in the 
leading order
\begin{eqnarray}
\lefteqn{
h_{\nu_{1}\nu_{2}\nu_{3}\nu_{4}}^{
(\sigma_{1}\sigma_{2}\lambda\sigma_{4},\mathrm{dir})} = 
\frac{e^2}{4\pi\epsilon_0\epsilon_s}
\sum_{ss'} \frac{1}{V^{2}}
} 
\\ &&  \times 
\int d^{3}r \int' d^{3} r'
\psi_{\nu_1}^{*}(\rr')\psi_{\nu_2}^{*}(\rr)
\nabla\frac{1}{|\rr-\rr'|}
\psi_{\nu_3}(\rr) \psi_{\nu_4}(\rr')
\nonumber \\ && \times 
 \int d^{3}\xi' u^{*}_{\sigma_{1}}(\xx',s')
u_{\sigma_{4}}(\xx',s')
\int d^{3}\xi u^{*}_{\sigma_{2}}(\xx,s)
\xx u_{\lambda}(\xx,s),
\nonumber
\label{matel2}
\end{eqnarray}
where $V$ is the volume of the unit cell. The zeroth order term as well 
as the term containing $\xx'$ vanish because the functions $u_
{\sigma_{1}}$ and $u_{\sigma_{4}}$ both have $s$-type symmetry. The 
integration over $\rr'$ can be extended onto the whole space because the 
singularity of $\nabla(1/r)$ is integrable in three dimensions.
Using 
the orthogonality of Bloch functions and the definition
\begin{equation}\label{bloch-dipol}
\sum_{s}\int d^{3}\xi u^{*}_{\sigma_{2}}(\xx,s)
\xx u_{\lambda}(\xx,s)=V\bm{a}_{\sigma_{2}\lambda},
\end{equation}
one then finds
\begin{eqnarray}
h_{\nu_{1}\nu_{2}\nu_{3}\nu_{4}}^{
(\sigma_{1}\sigma_{2}\lambda\sigma_{4},\mathrm{dir})} & = &
-\frac{i}{(2\pi)^{3}}\frac{e^2}{\epsilon_0\epsilon_s}
\int d^{3}q 
\frac{\qq\cdot\bm{a}_{\sigma_{2}\lambda}}{q^{2}}
\label{matel-gen} \\ && \times
\mathcal{F}_{\nu_{1}\nu_{4}}(\qq) 
\mathcal{F}^{*}_{\nu_{3}\nu_{2}}(\qq)
\delta_{\sigma_{1}\sigma_{4}},
\nonumber
\end{eqnarray}
where we used the identity
\begin{displaymath}
\nabla\frac{1}{|\bm{r}-\bm{r}'|}=-\frac{i}{2\pi^{2}}\int d^{3}q
\frac{\qq}{q^{2}}e^{i\qq\cdot(\bm{r}'-\bm{r})},
\end{displaymath}
and the form-factors are defined as
\begin{equation}
\mathcal{F}_{\nu\nu'}(\qq)=
\int d^{3}r \psi_{\nu}^{*}(\rr)e^{i\qq\cdot\rr}\psi_{\nu'}(\rr).
\label{Form-Factors}
\end{equation}
Note that, according to Eq.~\eqref{matel-gen}, the leading order term
in the matrix element responsible for the MEG (impact ionization)
process involves $\nabla (\rr-\rr')^{-1}$ (as opposed to just
$(\rr-\rr')^{-1}$ in an intraband matrix element), hence the resulting
quantity is proportional to $1/R^{2}$. This is a consequence of the
orthogonality of the Bloch functions that leads to the appearance of
the bulk interband matrix element 
of the position operator, $\bm{a}_{\sigma_{2}\lambda}$,
(which obviously has the dimension of length) in the interband Coulomb
term. Note 
that the resulting selection rules for the impact ionization process 
are the same as for the dipole-allowed optical transitions. 

In a crystal with zinc-blende structure, the topmost valence band
corresponds to the 3/2 band angular momentum. The non-zero matrix
elements $\bm{a}_{\sigma\lambda}$ between the four valence bands with
$\lambda=\pm 1/2, \pm 3/2$ and the  two conduction bands with
$\sigma=\pm 1/2$  are 
\begin{subequations}
\begin{eqnarray}
\bm{a}_{\pm \frac{1}{2},\pm\frac{3}{2}}=
\sqrt{3}\, \bm{a}_{\mp \frac{1}{2},\pm\frac{1}{2}} & = & 
\frac{a_{0}}{\sqrt{2}}
\left( \begin{array}{c} \mp 1 \\ -i \\ 0 \end{array} \right),
\label{axy} \\
\bm{a}_{\pm \frac{1}{2},\pm\frac{1}{2}} & = & 
a_{0}\sqrt{\frac{2}{3}}
\left( \begin{array}{c} 0 \\ 0 \\ 1 \end{array} \right),
\label{az}
\end{eqnarray}
\end{subequations}
where $a_{0}$ is a bulk material constant. Hence, we find
\begin{equation}\label{qa}
\qq \cdot \bm{a}_{\sigma_{2},\lambda} =
\alpha_{\sigma_{2}\lambda} qa_{0}
\sqrt{\frac{4 \pi}{3}} Y_{1,\Delta m}(\vartheta, \varphi),
\end{equation}
where $\Delta m=\lambda-\sigma_{2}$ and the non-zero coefficients are 
\begin{displaymath}
\alpha_{\pm\frac{1}{2},\pm\frac{3}{2}}=1,\quad
\alpha_{\mp\frac{1}{2},\pm\frac{1}{2}}=\frac{1}{\sqrt{3}},\quad
\alpha_{\pm\frac{1}{2},\pm\frac{1}{2}}=\sqrt{\frac{2}{3}}.
\end{displaymath}

\subsection{Surface polarization contribution}

The surface polarization term is smooth, therefore separation into 
short-range and long-range parts is not necessary. Again, we expand 
the potential into the Taylor series in $\xx,\xx'$,
\begin{eqnarray*}
U_{\mathrm{pol}}(\rr_{i}+\xx,\rr_{j}+\xx') & \approx &
U_{\mathrm{pol}}(\rr_{i},\rr_{j})+
\xx\cdot\nabla_{i}U_{\mathrm{pol}}(\rr_{i},\rr_{j}) \\ 
& + &\xx'\cdot\nabla_{j}U_{\mathrm{pol}}(\rr_{i},\rr_{j})+\ldots,
\end{eqnarray*}
and keep only the lowest order non-vanishing term. The first and third 
terms in the above expression lead to vanishing integrals in 
Eq.~\eqref{matel1} because of the orthogonality of the Bloch functions 
for different bands. Using the properties of Legendre polynomials and 
the addition theorem for spherical harmonics\cite{jackson98} we find 
for the second term
\begin{eqnarray*}
\lefteqn{\xx\cdot\nabla_{i}U_{\mathrm{pol}}(\rr_{i},\rr_{j})=
\frac{e^2}{\epsilon_0\epsilon_sR}\sum_{k}
\frac{r_{i}^{k-1}r_{j}^{k}}{R^{2k}}} 
\\ && \times
\left[ 
\frac{\xx\cdot\rr_{i}}{r_{i}}
\sum_{\substack{l<k\\ l-k\,\mathrm{even}}}\sum_{m=-l}^{l}
Y_{lm}^{*}(\theta_{i},\phi_{i})Y_{lm}(\theta_{j},\phi_{j})
\right. \\ &&
\left.
-\frac{\xx\cdot\rr_{j}}{r_{j}}
\sum_{\substack{l<k\\ l-k\,\mathrm{odd}}}\sum_{m=-l}^{l}
Y_{lm}^{*}(\theta_{i},\phi_{i})Y_{lm}(\theta_{j},\phi_{j})
 \right],
\end{eqnarray*}
where $(r_{i},\theta_{i},\phi_{i})$ are the spherical coordinates of 
the vector $\rr_{i}$. One substitutes the resulting expression into 
Eq.~\eqref{matel1}, performs the integration over the Bloch functions 
according to Eq.~\eqref{bloch-dipol} and changes the summation over 
$i,j$ into integration as previously. As a result, the surface 
polarization contribution to the matrix element is
\begin{displaymath}
h_{\nu_{1}\nu_{2}\nu_{3}\nu_{4}}^{(\sigma_{1}\sigma_{2}\lambda\sigma_{4},\mathrm{pol})}=
h_{\nu_{1}\nu_{2}\nu_{3}\nu_{4}}^{(\sigma_{1}\sigma_{2}\lambda\sigma_{4},\mathrm{pol-1})}+
h_{\nu_{1}\nu_{2}\nu_{3}\nu_{4}}^{(\sigma_{1}\sigma_{2}\lambda\sigma_{4},\mathrm{pol-2})},
\end{displaymath}
where 
\begin{eqnarray}
\lefteqn{h_{\nu_{1}\nu_{2}\nu_{3}\nu_{4}}^{(\sigma_{1}\sigma_{2}\lambda\sigma_{4},\mathrm{pol-}j)} = }
\label{matel3}
\\ && (-1)^{j-1}\frac{e^2}{\epsilon_0\epsilon_sR}
\sum_{k}\chi_{k}\int d^{3}r_{1}\int d^{3}r_{2}
\psi_{\nu_{1}}^{*}(\rr_{2})
\psi_{\nu_{2}}^{*}(\rr_{1})
\nonumber \\ && \times
\frac{r_{1}^{k-1}{r_{2}}^{k}}{R^{2k}} 
\frac{\bm{a}_{\sigma_{2}\lambda}\cdot\rr_{j}}{r_{j}}
\sum_{\substack{l<k\\ l-k+j\,\mathrm{odd}}}\sum_{m=-l}^{l}
Y_{lm}^{*}(\theta_{1},\phi_{1})
\nonumber \\&& \times
Y_{lm}(\theta_{2},\phi_{2})
\psi_{\nu_{3}}(\rr_{1})
\psi_{\nu_{4}}(\rr_{2})
\delta_{\sigma_{1},\sigma_{4}}
\nonumber
\end{eqnarray}
for $j=1,2$.

As previously, using Eq.~\eqref{axy} and Eq.~\eqref{az}, one finds
\begin{equation}\label{ra}
\rr \cdot \bm{a}_{\sigma_{2},\lambda} =
\alpha_{\sigma_{2}\lambda} ra_{0}
\sqrt{\frac{4 \pi}{3}} Y_{1,\Delta m}(\theta, \phi).
\end{equation}
From Eq.~\eqref{ra}, again the general scaling of the matrix element
as $1/R^{2}$ follows.

\subsection{Coulomb coupling between two- and four-particle
  configurations} 

Assuming that the NC is spherical and neglecting the 
spin-orbit coupling, the total spins of electrons and holes are
separately good quantum numbers. The bright states involving the hole
in a state with $\pm 3/2$ 
angular momentum will be denoted by 
$\left|\alpha\beta \uparrow\Downarrow\right\rangle$ and 
$\left|\alpha\beta \downarrow\Uparrow\right\rangle$, 
where $\alpha$ and $\beta$ denote
the electron and hole states (representing the relevant sets of
quantum numbers) and the arrows represent the values of the projection of 
the band angular momentum of the electron and hole on the selected
quantization axis. The states with dark spin configurations as well as
the states involving a hole with $\pm 1/2$ band angular momentum
(represented by $\uparrow,\downarrow$) are denoted in an analogous way.
Note that, e.g., $\Uparrow$ denotes hole spin 
$+3/2$, which results from a transition from the $\lambda=-3/2$
valence band.

The four-particle states are labeled by the electron spin
configuration ($S,T_{\pm,0}$) for singlet and the three triplet
states, respectively and by the hole spin configuration. If only the
topmost valence band ($j=3/2$) is included than the two-hole
configurations can be classified in the following way, which is
convenient for our purpose: The states with both holes in $\pm 3/2$
or both holes in $\pm 1/2$ states are combined in singlet-like and
triplet-like configurations $S^{3/2},T_{\pm,0}^{3/2}$ and
$S^{1/2},T_{\pm,0}^{1/2}$. For instance, in terms of the hole creation
operators $\hat{h}_{\mu,\lambda}^{\dag}$, the two-hole state with
$S^{3/2}$ and 
$T_{0}^{3/2}$ spin configurations is 
$|\mu\mu'\Sigma_{\mathrm{h}}\rangle=
(\eta_{\mu\mu'}/\sqrt{2})
(\hat{h}_{\mu\Uparrow}^{\dag} \hat{h}_{\mu'\Downarrow}^{\dag}
\pm \hat{h}_{\mu'\Uparrow}^{\dag}
\hat{h}_{\mu\Downarrow}^{\dag})|0\rangle$, 
where $\mu\ge\mu'$ (equality allowed only in the singlet
configuration) and  $\eta_{\mu\mu'} = 1/\sqrt{2}$ for $\mu=\mu'$, and
$\eta_{\mu\mu'} = 1$ otherwise. 
While these spin configurations are not
total spin eigenstates, they have a definite
parity under particle permutation, implying also a definite (opposite)
parity of the orbital wave functions. Hence, the resulting states
automatically diagonalize the hole exchange interaction. The spin
configurations with one hole in a $\pm 3/2$ state and one hole in the
$\pm 1/2$ state are obtained simply by symmetrizing (S) or
antisymmetrizing (A) the corresponding two-hole states with respect to the
orbital wave functions. For instance, 
$|\mu\mu',S_{\uparrow\Downarrow}\rangle=
(\eta_{\mu\mu'}/\sqrt{2})
(\hat{h}_{\mu\uparrow}^{\dag} \hat{h}_{\mu'\Downarrow}^{\dag}
+ \hat{h}_{\mu'\uparrow}^{\dag}
\hat{h}_{\mu\Downarrow}^{\dag})|0\rangle$, $\mu\ge\mu'$, and
$|\mu\mu',A_{\uparrow\Downarrow}\rangle=
(\eta_{\mu\mu'}/\sqrt{2})
(\hat{h}_{\mu\uparrow}^{\dag} \hat{h}_{\mu'\Downarrow}^{\dag}
- \hat{h}_{\mu'\uparrow}^{\dag}
\hat{h}_{\mu\Downarrow}^{\dag})|0\rangle$, $\mu>\mu'$.
The four-particle (biexciton) states are then labeled by 
$\left|\nu\nu'\Sigma_{\mathrm{e}};\mu\mu'\Sigma_{\mathrm{h}}\right>$, 
where 
$\nu, \nu'$ denote electron states, 
$\mu, \mu'$ are hole states, 
and $\Sigma_{\mathrm{e}},\Sigma_{\mathrm{h}}$ 
represent spin configurations. 

The couplings between the two-particle and four-particle
configurations, which are responsible for the impact ionization
process, are then expressed in terms of the Coulomb matrix elements
given by Eqs.~\eqref{matel-gen} and \eqref{matel3}. For instance, the
non-zero couplings between the four-particle states with both holes in
the spin-$\pm 3/2$ states and the two-particle state
$|\alpha\beta\uparrow\Downarrow\rangle$ are (denoting
$h_{\Sigma_{\mathrm{e}}\Sigma_{\mathrm{h}}}= 
\left\langle \nu\nu'\Sigma_{\mathrm{e}},\mu\mu'\Sigma_{\mathrm{h}}
\left| H \right| \alpha\beta \Downarrow \right\rangle$)
\begin{eqnarray*}
h_{SS^{3/2}} & = & 
\frac{\eta_{\nu\nu'}\eta_{\mu\mu'}}{2} \left[
\delta_{\mu\beta}\left(  h_{1} +h_{2} \right)
+\delta_{\mu'\beta}\left(  h_{3} +h_{4} \right)\right], \\
h_{ST_{0}^{3/2}}  & = & 
\frac{\eta_{\nu\nu'}}{2} \left[
\delta_{\mu\beta}\left( -h_{1} -h_{2} \right)
+\delta_{\mu'\beta}\left(  h_{3} +h_{4} \right)\right], \\
h_{T_{0}S^{3/2}}  & = & 
\frac{\eta_{\mu\mu'}}{2} \left[
\delta_{\mu\beta}\left(  h_{1} -h_{2} \right)
+\delta_{\mu'\beta}\left(  h_{3} -h_{4} \right)\right], \\
h_{T_{0}T_{0}^{3/2}} & = & 
\frac{1}{2} \left[ 
\delta_{\mu\beta}\left( -h_{1} +h_{2} \right)
+\delta_{\mu'\beta}\left(  h_{3} -h_{4} \right)\right], \\
h_{T_{+}T_{-}^{3/2}} & = & 
\delta_{\mu\beta}\left( -h'_{1} +h'_{2} \right)
+\delta_{\mu'\beta}\left(  h'_{3} -h'_{4} \right),
\end{eqnarray*}
where 
$h_{1}=h_{\nu\nu'\mu'\alpha}^{\left( \uparrow\downarrow\Uparrow\uparrow \right)}$,
$h_{2}=h_{\nu'\nu\mu'\alpha}^{\left( \uparrow\downarrow\Uparrow\uparrow \right)}$,
$h_{3}=h_{\nu\nu'\mu\alpha}^{\left( \uparrow\downarrow\Uparrow\uparrow \right)}$,
$h_{4}=h_{\nu'\nu\mu\alpha}^{\left( \uparrow\downarrow\Uparrow\uparrow \right)}$,
$h'_{1}=h_{\nu\nu'\mu'\alpha}^{\left( \uparrow\uparrow\Downarrow\uparrow \right)}$,
$h'_{2}=h_{\nu'\nu\mu'\alpha}^{\left( \uparrow\uparrow\Downarrow\uparrow \right)}$,
$h'_{3}=h_{\nu\nu'\mu\alpha}^{\left( \uparrow\uparrow\Downarrow\uparrow \right)}$,
$h'_{4}=h_{\nu'\nu\mu\alpha}^{\left( \uparrow\uparrow\Downarrow\uparrow \right)}$.
Since one of the holes is a spectator in the impact ionization process and its
state is conserved there is no coupling between this two-particle
state and any state with both holes in a $\pm 1/2$ spin state.
The results for the state $\left|\alpha\beta\downarrow\Uparrow\right>$
are obtained by flipping all  the spins and the couplings for a
two-particle state with a hole in a $\pm 1/2$ state are easily derived
by exchanging the role of the $\pm 3/2$ and $\pm 1/2$ hole spins.

With similar notation as above and $\lambda=\uparrow,\downarrow$, the
non-vanishing couplings between 
four-particle states with one hole in a $\pm 1/2$ state and one in a
$\pm 3/2$ state and the same two-particle state are 
\begin{eqnarray*}
h_{SS_{\lambda\Downarrow}} & = & 
\frac{\eta_{\nu\nu'}\eta_{\mu\mu'}}{2} \left[
\delta_{\mu\beta}\left(  h_{1}'' +h_{2}'' \right)
+\delta_{\mu'\beta}\left(  h_{3}'' +h_{4}'' \right)\right], \\
h_{SA_{\lambda\Downarrow}} & = & 
\frac{\eta_{\nu\nu'}}{2} \left[
-\delta_{\mu\beta}\left(  h_{1}'' +h_{2}'' \right)
+\delta_{\mu'\beta}\left(  h_{3}'' +h_{4}'' \right)\right], \\
h_{T_{0}S_{\lambda\Downarrow}} & = & 
\frac{\eta_{\mu\mu'}}{2} \left[
\delta_{\mu\beta}\left(  h_{1}'' -h_{2}'' \right)
+\delta_{\mu'\beta}\left(  h_{3}'' -h_{4}'' \right)\right], \\
h_{T_{0}A_{\lambda\Downarrow}} & = & 
\frac{1}{2} \left[
\delta_{\mu\beta}\left( - h_{1}'' +h_{2}'' \right)
+\delta_{\mu'\beta}\left(  h_{3}'' -h_{4}'' \right)\right], \\
h_{T_{+}S_{\lambda\Downarrow}} & = & 
\frac{\eta_{\mu\mu'}}{\sqrt{2}} \left[
\delta_{\mu\beta}\left(  h_{1}''' -h_{2}''' \right)
+\delta_{\mu'\beta}\left(  h_{3}''' -h_{4}''' \right)\right], \\
h_{T_{+}A_{\lambda\Downarrow}} & = & 
\frac{1}{\sqrt{2}} \left[
\delta_{\mu\beta}\left( - h_{1}''' +h_{2}''' \right)
+\delta_{\mu'\beta}\left(  h_{3}''' -h_{4}''' \right)\right],
\end{eqnarray*}
where $h_{i}''$ and $h_{i}'''$ are defined as $h_{i}$ and $h_{i}'$,
respectively, but with the third spin (upper) index replaced by
$\lambda$. Again, the results for the two-particle state
$\left|\alpha\beta\downarrow\Uparrow\right>$ 
are obtained by flipping all  the spins.

Thus, we have characterized the Coulomb couplings between
two-particle and four-particle configurations within the envelope
function approach. The matrix elements for transitions between various
subbands can be used in a calculation of couplings in an arbitrary
envelope function model, including the common one based on an 8-band
$k\cdot p$ Hamiltonian (obviously, the classification of the
four-particle configuration must then be extended to account or the
spin-orbit coupling). This rather complex task is beyond the scope of
this paper. In the following section, we limit ourselves to the
simplest application of the results obtained above to a single-band
model. 

\section{Matrix element for a simple model 
of nanocrystal wave functions}
\label{sec:simple-model}

In this section, as the simplest example of an application of the
general envelope function formalism of Sec.~\ref{sec:model}, 
we find the Coulomb matrix elements between the 
2- and 4-particle states assuming simple single-band carrier 
wave functions. While perhaps not quantitatively accurate, this calculation
yields a useful estimate of the overall magnitude of the couplings and
of the resulting degree of mixing between two- and four-particle
configurations as well as some statistics of the coupling strengths
that, taken globally, may be close to the actual ones. Below, we
define the simple model of wave functions which is then used to
implement the general findings of the previous section.

\subsection{Model of the wave functions}

An InAs NC is modeled as a spherical potential well of radius 
$R$ with infinite potential walls. The envelope wave function is given 
by
\begin{equation}
\psi_{\nu}(\bm{r})=
\frac{1}{R^{3/2}}N_{nl}Y_{lm}(\theta,\phi) j_{l}(x_{ln}r/R),
\label{envelope}
\end{equation}
where we write explicitly $\nu=(nlm)$, $x_{ln}$ is the $n$th zero of 
the spherical Bessel function $j_{l}$ and 
$N_{nl}=\sqrt{2}/|j_{l+1}(x_{ln})|$. In our model, we neglect band 
mixing and include the Coulomb interaction between electrons and holes 
in the lowest order only.

The corresponding energy levels for electrons and holes are 
$E_{nl}^{(\mathrm{e,h})}=\hbar^{2}x_{ln}^{2}/( 2m^{*}_{\mathrm{e,h}}R^2 ).$
We take the heavy hole effective mass $m^{*}_{\mathrm{h}}=0.35m_{0}$, 
where $m_{0}$ is the free electron mass, and use the implicit formula 
for the energy-dependent electron mass in the decoupled bands 
approximation \cite{efros98} 
$m^{*}_{\mathrm{e}}=m_{0}[\alpha
+E_{\mathrm{P}}/(E_{\mathrm{g}}+E_{nl})]^{-1}$,
where $E_{\mathrm{P}}=22.2$~eV, $E_{\mathrm{g}}=0.418$~eV is the bulk 
band gap, and the parameter $\alpha=0.77$ accounts for the coupling to 
higher bands. Using the implicit formula for the electron and a 
constant effective mass for the hole is motivated by the small 
effective mass of the former which leads to larger kinetic energies as 
compared to the hole.

In the energy levels of the few-particle states, we include the lowest 
order corrections due to Coulomb interactions, including the surface 
polarization terms\cite{Brus84}. In the energies of the four-particle 
states also exchange interactions are taken into account.

\subsection{Direct Coulomb coupling}
In Eq.~\eqref{Form-Factors}, we substitute the wave functions from 
Eq.~\eqref{envelope} and use the expansion \cite{jackson98} 
\begin{displaymath}
e^{i\qq\cdot\rr} =
4\pi \sum_{lm} i^{l} j_{l}(qr) 
Y_{lm}^{*}(\theta,\phi)Y_{lm}(\vartheta,\varphi),
\end{displaymath}
where $\left( r, \theta, \phi \right)$ and $\left( q,
  \vartheta,\varphi \right)$ are the spherical coordinates of the
vectors $\rr$ and $\qq$ respectively. As a result we get 
\begin{eqnarray}
\mathcal{F}_{\nu\nu'}(\qq) & = &
4\pi\left( -1 \right)^{m-m'} 
\sum_{l''=\left| l-l' \right|}^{l+l'}
i^{l''}f_{ll''l'}^{nn'} \left( qR \right) 
\nonumber \\ && \times
G_{ll''l}^{m,m-m',m'} Y_{l'',m-m'}(\vartheta,\varphi), 
\label{FormFaktor}
\end{eqnarray}
where
\begin{displaymath}
f_{ll''l'}^{nn'}(u)=
 N_{lm}N_{l'm'}\int_{0}^{1}dy y^2 j_{l''}(uy)j_{l}(x_{ln}y)j_{l'}(x_{l'n'}y)
\end{displaymath}
and 
\begin{eqnarray}
G_{ll'l''}^{mm'm''} & = & 
\int_{0}^{2\pi}d\phi\int_{0}^{\pi}d\theta\sin{\theta}
\label{Gaunt} \\ 
&& \times 
Y_{lm}^{*}(\theta,\phi)
Y_{l'm'}(\theta,\phi)
Y_{l''m''}(\theta,\phi)
\nonumber
\end{eqnarray}
are Gaunt coefficients. 

Using Eqs.~\eqref{matel-gen}, \eqref{qa}, and~\eqref{FormFaktor}
we have
\begin{eqnarray*}
\lefteqn{
h_{\nu_{1}\nu_{2}\nu_{3}\nu_{4}}^
{(\sigma_{1}\sigma_{2}\lambda\sigma_{4},\mathrm{dir})} = 
} \\ && 
-\frac{4e^{2}a_{0}}{\sqrt{3\pi} \epsilon_{0}\epsilon_{\mathrm{s}}R^{2}}
\alpha_{\sigma_{2}\lambda} \delta_{\sigma_{1},\sigma_{4}}
\\ && \times
\sum_{l=|l_{1}-l_{4}|}^{l_{1}+l_{4}} 
\sum_{l'=|l_{2}-l_{3}|}^{l_{2}+l_{3}} i^{l-l'+1}
\int_{0}^{\infty} duu  f_{l_{1}ll_{4}}^{n_{1}n_{4}}(u)
f_{l_{3}l'l_{2}}^{n_{3}n_{2}}(u)
\\ && \times
G_{l_{1},l,l_{4}}^{m_{1},m_{1}-m_{4},m_{4}}
G_{l_{3},l',l_{2}}^{m_{3},m_{3}-m_{2},m_{2}}
G_{l,1,l'}^{m_{1}-m_{4},\pm 1,m_{3}-m_{2}},
\end{eqnarray*}
where $\Delta m=\lambda-\sigma_{2}$.
The $1/R^{2}$ dependence of the matrix element is explicit in this result.
Note also that $l-l'$ must be odd for the Gaunt coefficients to be non-zero
so that the matrix elements are real.

\subsection{Surface polarization contribution}
In the surface polarization-related term [Eq.~\eqref{matel3}], we use
Eq.~\eqref{ra}
and substitute the wave functions from Eq.~\eqref{envelope}. To reduce 
the product of four harmonics, we expand one pair into the Gaunt series,
\begin{eqnarray*}
\lefteqn{Y_{lm}^{*}\left( \theta,\phi\right)
Y_{1,\Delta m}\left( \theta,\phi\right) =} \\ 
&&\left( -1 \right)^{m}\sum_{l'=| l-1 |}^{l+1}
G_{l'1l}^{\Delta m-m,\Delta m,-m}Y_{l',\Delta m-m}(\theta,\phi).
\end{eqnarray*}
The resulting integrals are performed using Eq.~\eqref{Gaunt}. As a 
result, one finds
\begin{eqnarray*}
\lefteqn{h_{\nu_{1}\nu_{2}\nu_{3}\nu_{4}}^{
(\sigma_{1}\sigma_{2}\lambda\sigma_{4},\mathrm{pol}-1)} =
\sqrt{\frac{4\pi}{3}}\frac{e^{2}a_{0}}{\epsilon_{0}\epsilon_{\mathrm{s}}R^{2}}
\alpha_{\sigma_{2}\lambda} \delta_{\sigma_{1}\sigma_{4}}}
\\ && 
 \times
\sum_{k} \chi_{k} 
C_{k+1}^{\left( n_{2}l_{2}\right) \left( n_{3}l_{3}\right)}
C_{k+2}^{\left( n_{1}l_{1}\right) \left( n_{4}l_{4}\right)}
\sum_{\substack{l<k\\ l+k\,\mathrm{even}}} 
\sum_{l'=|l-1|}^{l+1}
\\ && 
\times G_{l_{1}ll_{4}}^{m_{1},m_{1}-m_{4},m_{4}}
G_{l_{3}l'l_{2}}^{m_{3},m_{3}-m_{2},m_{2}}
G_{l'1l}^{m_{1}-m_{4},\Delta m,m_{3}-m_{2}},
\end{eqnarray*}
and
\begin{eqnarray*}
\lefteqn{
h_{\nu_{1}\nu_{2}\nu_{3}\nu_{4}}^{(\sigma_{1}\sigma_{2}\lambda\sigma_{4},\mathrm{pol}-2)} =
-\sqrt{\frac{4\pi}{3}}\frac{e^{2}a_{0}}{\epsilon_{0}\epsilon_{\mathrm{s}}R^{2}
} \alpha_{\sigma_{2}\lambda} \delta_{\sigma_{1}\sigma_{4}}} \\ 
&& \times
\sum_{k} \chi_{k} 
C_{k+1}^{\left( n_{2}l_{2}\right) \left( n_{3}l_{3}\right)}
C_{k+2}^{\left( n_{1}l_{1}\right) \left( n_{4}l_{4}\right)}
\sum_{\substack{l<k\\ l+k\,\mathrm{odd}}} 
\sum_{l'=|l-1|}^{l+1}
\\ && \times
G_{l_{1}l'l_{4}}^{m_{1},m_{1}-m_{4},m_{4}}
G_{l_{3}ll_{2}}^{m_{3},m_{3}-m_{2},-m_{2}}
G_{l'1l}^{m_{1}-m_{4},\Delta m,m_{3}-m_{2}}, 
\end{eqnarray*}
where
\begin{displaymath}
C_{k}^{\left( nl\right) \left( n'l'\right)} =
N_{nl}N_{n'l'}
\int_{0}^{1} dxj_{l} \left( x_{nl}x \right) x^{k} j_{l'} \left( x_{n'l'}x \right)
\end{displaymath}
and $\Delta m$ is defined as previously. Again, these results show an
explicit $1/R^{2}$ dependence.

\section{Results}
\label{sec:results}

\begin{figure}[!h]
\includegraphics[width=84mm]{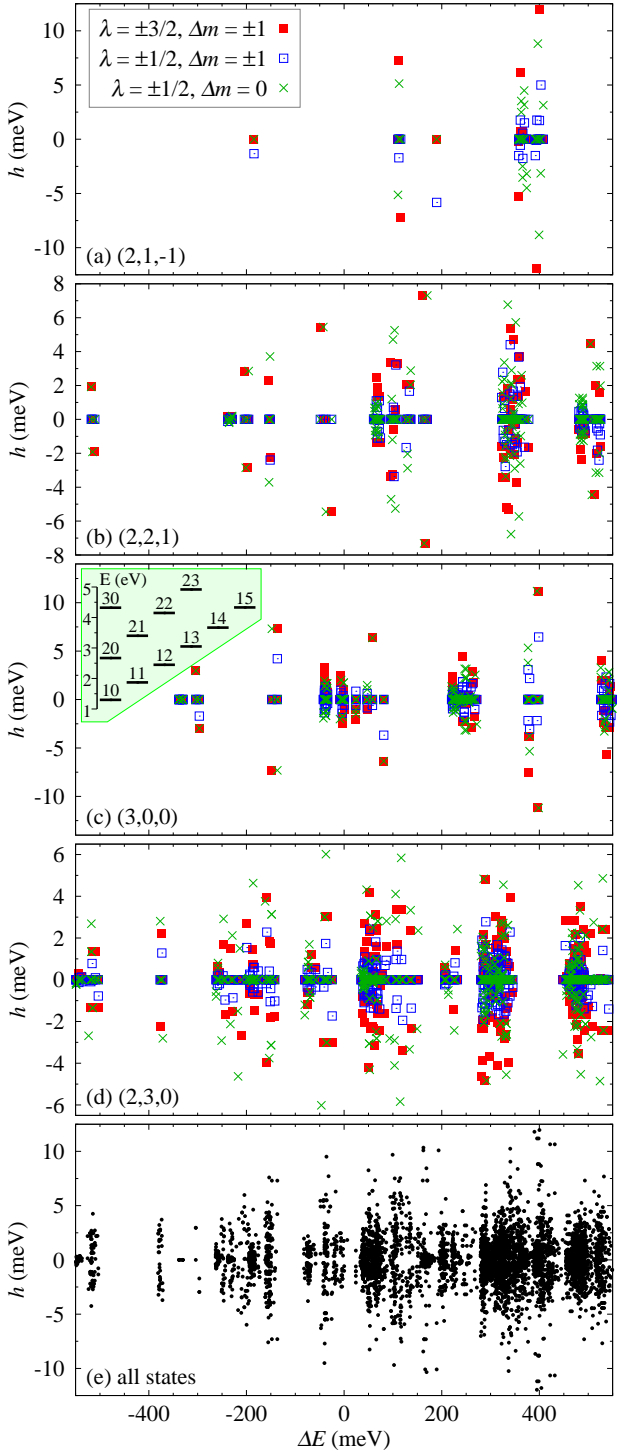}
\caption{(a-c) The matrix elements $h$ between X and BX
  states vs. the energy distance between these states for
  selected X states, as shown in the labels, for a NC with
  the radius $R=3$~nm. The point style encodes 
  three groups of spin configurations of the BX state as
  shown in the panel (a) (see also the text). (d) The
  matrix elements and X-BX energy distances for all the X sates below 
 5~eV.  Inset in (c): the spectrum of the X states 
  (the digits show the $n$ and $l$ values).} 
\label{fig:h-DE}
\end{figure} 
 
In this section, we present results of calculations performed
within the simple single-band model presented above. We focus on the
general statistical distribution of the coupling strengths between
optically active (bright) single-exciton (X) and biexciton (BX) states vs.
the energy differences between the two coupled states. This allows us 
to estimate the degree of mixing between the bright X and
BX states. The calculations are performed for a single spherical
NC with the radius $R=R_{0}=3$~nm or for a slightly
inhomogeneous ensemble of NCs with the sizes given by the
Gaussian distribution of their radii with the mean $R_{0}$
and the full width at half maximum (FWHM) of $\sigma=0.3$~nm. 
We take $a_{0}=1.9$~nm, as estimated from the InAs parameters
routinely used in \mbox{$\bm{k}\cdot\bm{p}$} calculations \cite{yu05}.
For definiteness, we focus on 
the bright X states with the hole with spin projection $+3/2$ that lie
in the energy range below 5~eV. This set contains 53 states in  a
NC of $3$~nm radius (the spectrum is showed in the
inset in Fig.~\ref{fig:h-DE}(c)). The lowest BX state, which sets the
energetical onset of the MEG process, is at 2.6~eV.

In order to characterize the typical X-BX Coulomb coupling strengths
and the distribution of energies of the coupled BX
configurations, in Fig.~\ref{fig:h-DE}(a-c) we graphically represent
these couplings for three selected bright X states (indicated by
the values of the quantum numbers $(n,l,m)$, identical for the electron
and the hole, shown in the label of each panel). Couplings to BX states
within the energy interval of $\pm 550$~meV around the energy of a
given X state are shown. Each symbol
corresponds to a single BX state coupled to a given
X state and its position shows the energy distance from
the X state and the value of the Coulomb
matrix element coupling the X and BX states. The BX states are divided
into three groups according to their spin configurations: 5
configurations with the spin of the newly created hole 
$\lambda=\pm 3/2$, 6 configurations with $\lambda=\pm 1/2$ and the
envelope angular
momentum change $\Delta m=\pm 1$, and 6 configurations with
$\lambda=\pm 1/2$ 
and $\Delta m=\pm 1$. These three groups are coded into the symbol
styles, as shown in the key inserted in  Fig.~\ref{fig:h-DE}(a).

One can see that typical values of the X-BX couplings are up to
several meV but most of them are at most on the order of 1~meV (we
have found a small number of stronger couplings, even over 30~meV, but
only between energetically very distant states). There is a clear
pattern in the spectral distribution of the coupled BX states which
results from the shell structure of the NC spectrum. An
important feature is the growing number of the coupled BX states which
is consistent with the rapid growth of the overall density of BX
states with increasing energy.

The features observed in the case of the three selected X states
shown in Fig.~\ref{fig:h-DE}(a-c) are confirmed by the analysis of the
combined distribution for all the X states with energies below 5~eV,
shown in Fig.~\ref{fig:h-DE}(d). Altogether, there are almost 7200 BX states
coupled to the 53 X states in this energy range, which is still only a
tiny fraction of the total number of the BX states in this energy
interval. 

\begin{figure}[tb]
\includegraphics[width=8.5cm]{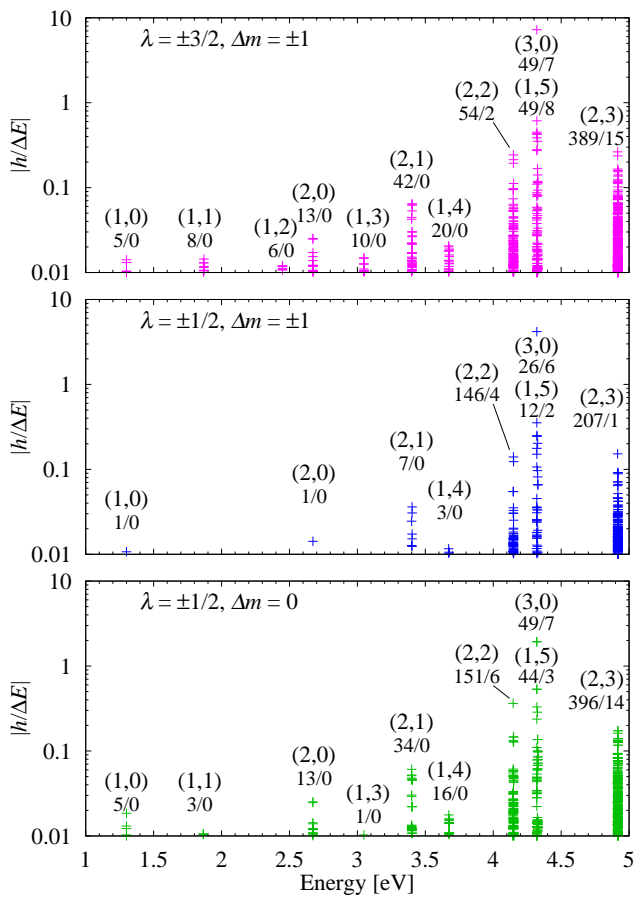}
\caption{The ratio of the X-BX matrix element $h$ to the energy
  distance $\Delta E$
  between the coupled states plotted as a function of the energy of
  the X state involved for a NC with the radius $R=3$~nm. 
 The label $(n,l)$ indicates the quantum  numbers
  and the label $a/b$ shows the numbers of coupled BX states with
  $|h/\Delta E|>0.01$ ($a$) and  $|h/\Delta E|>0.01$ ($b$).}
\label{fig:zbiorcze}
\end{figure}

The results discussed above suggest that the growing number of remote BX states 
that are sufficiently strongly coupled to a given X state can
considerably contribute to the mixing between X and BX states. Hence, a
picture based on BX states in vicinity of the X state (an ``energy
window'') may be 
misleading. In order to achieve a more complete picture we have found
all the states in the energy range $\pm 4$~eV from each of the X
states under consideration. This allows us to account for all the BX states for
which the ratio of the X-BX matrix element $h$ to the energy
distance $\Delta E$ is greater than 0.01, except for the uncommon
cases of $|h|>40$~meV (we have not found a single instance of such a
large coupling). The ratio $|h/\Delta E|$ is an important parameter
as it determines (within the range of applicability of the
perturbation theory) the admixture of the BX state into the X state
which is crucial for the efficiency of the MEG process. In
Fig.~\ref{fig:zbiorcze}, we plot this ratio as a function of the
energy of the X state involved. For more clarity, the results are
divided into three groups according to the spin configurations of the
BX state, as previously. Each ``stack'' of points in this figure
corresponds to a set of states with fixed values of the quantum
numbers $n$ and $l$, as denoted in the figure (with the exception of
the states $(3,0,0)$ and $(1,5,m)$ which are accidentally almost
degenerate for this NC size). 
As expected, the number of coupled BX states tends to grow with the
energy of the X state which can again be attributed to the growing density
of states of the BX states. As one can see, in spite of the enormous
number of BX states in the energy range taken into account the number
of states with $|h/\Delta E|>0.01$ can only reach several tens for a
single X state (note that the numbers in
the figure give the total number of BX states coupled to all the
$2l+1$ X states). Moreover, only for a small fraction of these states
one finds $|h/\Delta E|>0.1$ (which corresponds to an admixture of BX
state above 1\%). For the specific NC size unser study only
in one case a close resonance between an X state and a coupled BX
state was found that resulted in the $|h/\Delta E|$ ratio exceeding
1.

\begin{figure}[tb]
\includegraphics[width=85mm]{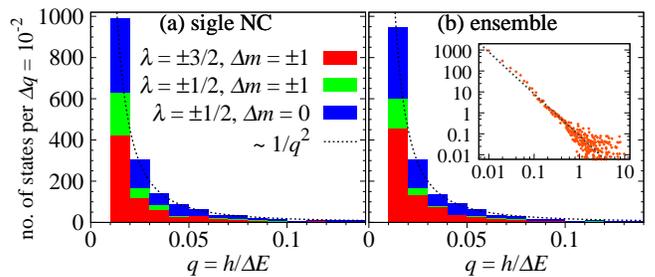}
\caption{Histograms of the number of coupled BX states
as a function of the ratio $q=|h/\Delta E|$. The color coding represents
three groups of spin configurations of the BX states and the dotted
lines show a $1/q^{2}$ dependence. (a) A single NC with
$R=3$~nm. (b) An ensemble of NCs with the mean radius
$R_{0}=3$~nm and the FWHM of the radius distribution of $0.3$~nm.}
\label{fig:histogram-1}
\end{figure}

Comparing the small number of states with $|h/\Delta E|>0.01$ found in the
$\pm 0.55$~eV energy window (as shown in Fig.~\ref{fig:h-DE}) with the
much larger numbers found in the broad energy range
(Fig.~\ref{fig:zbiorcze}) one might conclude that remote states
play an important role in the X-BX state mixing and, in
consequence, in the MEG efficiency. From the point of view of
theoretical modeling, this opens the critical question whether the
actual situation can be reasonably approximated by a model that takes
into account only BX states in a certain, sufficiently large spectral
window around a given X state or, in other words, whether the 
approximate results converge sufficiently fast when extending the
spectral window. In order to approach the answer to this question we have
estimated the statistical distribution of the values of $q=|h/\Delta E|$
based on our simplified NC model. The result for a single
NC with $R=3$~nm is shown in 
Fig.~\ref{fig:histogram-1}(a).  In order to obtain a single
characteristics we present a joint distribution of the $|h/\Delta E|$
ratios for all the X states below 5~eV, divided again in the three
groups of spin configurations. As could be expected, the number of
weakly coupled BX states is the highest, while the number of states
coupled by larger matrix elements decreases quickly with the growing
strength of the coupling (note that the number of BX states in the lowest
sector of $q$, not shown in the figure, is formally infinite). The
same tendency is seen in 
Fig.~\ref{fig:histogram-1}(b) where an analogous distribution is shown
in the same way for an  ensemble of NC in which we have
assumed a Gaussian 
distribution of the radii with the mean $R_{0}=3$~nm and the FWHM
equal to $0.3$~nm. In both cases, a dependence 
of the form $1/q^{2}$ is found, which is rigorously confirmed by the
logarithmic plot in the inset to Fig.~\ref{fig:histogram-1}(b), where
this power law dependence is seen to be maintained over a
surprisingly broad range of the $q$ values. 

The fact that the exponent of this power law distribution is equal to 2
is remarkable: According to the perturbation theory (which is valid
for small values of $q$), the admixture of a single BX state to a given
X state is equal to $q^{2}$. Hence, if there are $N_{q}$ states
with a certain value of $q$ then their joint contribution is equal to
$N_{q}q^{2}$. The scaling $N_{q}\sim 1/q^{2 }$ means that the growing
number of weakly coupling states exactly compensates the decreasing
magnitude of the matrix element so that, on the
average, BX states with all the coupling strengths contribute
equally. This has the immediate consequence that discarding the
part of the coupled states with $q<q_{\mathrm{min}}$ generates a
computational error that decreases proportionally to
$q_{\mathrm{min}}$. This means that, in principle, a numerical
computation can be performed with arbitrary accuracy based on a
properly selected set of BX states. In fact, since the values of the
matrix elements seem to be bounded, this implies convergence in terms
of the width of the energy window assumed. This property, that holds
only for a power law exponent below 3, is very desirable and usually
implicitly taken for granted in numerical computations 
\cite{franceschetti06,rabani08,allan08,rabani10,baer12,korkusinski11}
but it seems by no means obvious.

\begin{figure}[tb]
\includegraphics[width=85mm]{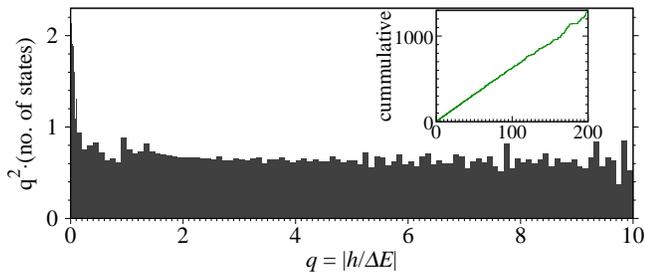}
\caption{Histogram of the number of coupled BX states multiplied by
  $q^{2}$ for an ensemble of NCs with the mean radius
$R_{0}=3$~nm and the FWHM of the radius distribution of $0.3$~nm.}
\label{fig:histogram-2}
\end{figure}

A direct confirmation of this ``homogeneous contribution'' property
resulting from the $1/q^{2}$ scaling is presented in
Fig.~\ref{fig:histogram-2}. Here each histogram bar is multiplied by
$q^{2}$. As a result, a remarkably constant distribution is obtained,
which is particularly visible in the inset, where the cumulative
distribution is shown, which has a linear form across a very wide
range of values. While this flat distribution is an interesting
feature, it should be noted that the value of 
$q^{2}N_{q}$ has no direct physical meaning  at high $q$ where the
perturbation theory is not applicable. In fact, this part of the
distribution is due to very few BX states that come to resonance with
some of the X states for a certain value of $R$. At resonance, the
value of $q$ is infinite but the maximum admixture in the case of just
two resonant states is 1/2.  

\begin{figure}[tb]
\includegraphics[width=70mm]{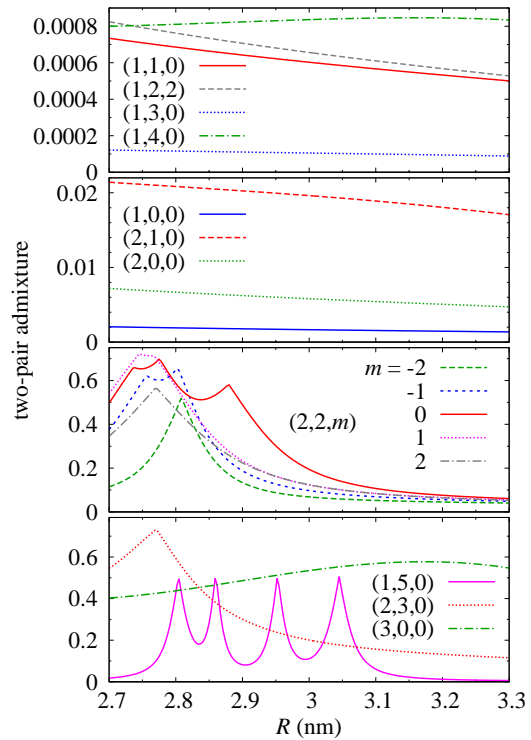}
\caption{Biexciton admixtures to selected exciton states.}
\label{fig:mix}
\end{figure}

Finally, let us estimate the degree of mixing between the X and
BX states. In Fig.~\ref{fig:mix}, we show the BX admixture to the
selected X
states considered here as a function of the NC radius. This
is obtained separately for each X state 
by diagonalizing the Hamiltonian including only the X state in
question and all the BX sates directly coupled to this X state by
matrix elements for which $q\ge 0.01$. For each $R$, the eigenstate
$\Psi_{0}$ with the highest $X$ contribution is found and the BX
admixture is determined as 
\begin{displaymath}
P_{\mathrm{BX}}=
\sum_{i}\left|\left\langle BX_{i}|\Psi_{0} \right\rangle \right|^{2} =
1-\left|\left\langle X|\Psi_{0} \right\rangle \right|^{2},
\end{displaymath}
where $|X\rangle$ is the X state and $|BX_{i}\rangle$ are all the
coupled BX states. We show the results for one selected state out of each
subspace with fixed $n$ and $l$, except for the states $(2,2,m)$ which
are presented in full to see that the admixture to states that differ
only by the value of $m$ is very similar (which is due to the fact
that these states are almost degenerate). 

As expected, the low energy states $(1,0,0)$ and $(1,1,m)$, lying
below the theoretical MEG threshold of 2.6~eV, have the BX admixtures well below
1\%. However, the same is true for states lying up to 1~eV above the
threshold. This property results from the selection rules that hold in
the MEG process, as discussed in Sec.~\ref{sec:model}: An X state
needs to be located in the vicinity of a BX state to which it is
coupled, which becomes likely only when the density of BX states
becomes sufficiently high. On the other hand, the states with energies
above 4~eV are typically much more strongly mixed with BX states. Even
in these case, in spite of hundreds of BX states coupled to each such
X state, the admixture only in some cases approaches 80\% for a
certain NC radius and for most (but not all) states drops
down as $R$ increases. Note that the admixture strongly depends on the
NC size and in the case of the state $(1,5,0)$ shows strong
oscillations. In general, while the X states are clearly not
completely dissolved in the densely distributed BX states, the degree
of mixing between X and BX states becomes considerable at energies a
few eV above the MEG threshold which may lead to the MEG
efficiency of a few tens per cent.

\section{Discussion, conclusions and outlook}
\label{sec: concl}

The main result presented in this paper is the method for calculating Coulomb
matrix elements between  exciton and biexciton states in semiconductor
nanocrystals based on the envelope function formalism. We have shown
that such a calculation requires proper treatment of the Bloch parts
of the carrier wave functions which, in the leading order, leads to
spin selection rules identical to those holding for optical interband
transitions (however, the rules for envelope states are
different). Moreover, the resulting matrix elements are additionally 
scaled by the ratio of the lattice constant to the NC radius,
as compared to the usual (intraband) Coulomb couplings. As a result,
the Coulomb coupling between X and BX states scale as
$1/R^{2}$. 

Once the matrix elements between single-band states are known they can
be used for calculating X-BX couplings using more exact carrier
states found by diagonalizing the 8-band Kane (\mbox{$\bm{k}\cdot\bm{p}$})
Hamiltonian. This 
approach has been found to correctly reproduce the NC
spectrum down to very small  sizes \cite{banin98} and, when combined
with the results presented here, can provide quantitatively accurate
description of the MEG process in NCs. For more accuracy,
Coulomb correlations, in particular in BX states \cite{korkusinski11},
could also be included. While atomistic models may offer more
accurate single particle wave functions and allow one to include
more system features (e.g., surface defects \cite{jaeger12}), their
high computational costs limits the extent to which few-particle
states and couplings between them can be treated. Typically, when
following an atomistic approach, one is
forced to restrict the calculations to an energy window around a given
X or BX state and to truncate the basis of BX states when simulating
the system dynamics \cite{korkusinski11}. From this point of view, the
method based on the envelope function formalism may offer a complementary
approach to the trade-off between the accuracy of single-particle states
and the reliability of few-particle modeling in which the accuracy of
the atomistic models is sacrificed in favor of lowering the
computational effort, which offers considerably more flexibility on
the subsequent stages of theoretical analysis, including the system
dynamics \cite{rabani10,korkusinski11} and dissipative evolution
\cite{shabaev06,witzel10,piryatinski10,velizhanin11a}, where finding
the X-BX coupling is the essential prerequisite for further
modeling. An additional benefit is the transparent 
nature of our envelope function method, which offers mostly analytical
treatment and does not rely on large computational resources or
dedicated software, hence can easily be employed by a broad community
of researchers.

While performing full multi-band \mbox{$\bm{k}\cdot\bm{p}$}
calculations is beyond the
scope of this work, we have presented some preliminary estimates of
the statistical distribution of the coupling magnitudes and the energies
of the coupled states using a very simple single-band envelope
function approach. Such overall statistical conclusions are likely to
be valid even if the underlying characteristics of individual states
are not absolutely accurate. We have shown that the number of BX
states coupled to X states form a certain energy range scales as
inverse square of the ratio of the coupling magnitude to the energy
separation. This scaling property is remarkable as it guarantees that
the contribution of remote states is finite and controllable, which
justifies limiting the computation to an energy window (no matter what
computational method is chosen)
\cite{franceschetti06,rabani08,allan08,rabani10,baer12,korkusinski11}. 

Finally, we have estimated the degree of mixing between X and BX
states as a function of the NC size. Very small BX admixture to X
states has been found below and within 1~eV above the MEG
threshold. Much larger mixing, reaching 80\%, appears for
higher-energy states. The amount of BX admixture to this states varies
quite strongly when the NC radius is changed by a fraction of
a nanometer. This may suggest that modeling based on a single
NC size may not be representative for average properties of
an ensemble.

\acknowledgements

This work was supported in part by the TEAM programme of the
Foundation for Polish Science, co-financed from the European Regional
Development Fund.


\end{document}